\documentstyle[aps,multicol,epsf]{revtex} 
\def\dir{.}
\def\d{{\rm d}}
\begin{document}
\title{Fluctuation-Induced Interactions between Rods on a Membrane}
\author{Ramin Golestanian}
\address{Institute for Advanced Studies in Basic Sciences,
Zanjan   45195-159, Iran}
\author{Mark Goulian\footnote{Present address:
{\it Center for Studies in Physics and Biology,
The Rockefeller University, 1230 York Avenue,
New York, NY 10021.}}}
\address{Exxon Research and Engineering,
Annandale, NJ  08801}
\author{Mehran Kardar}
\address{Department of Physics, Massachusetts Institute of
Technology, Cambridge, MA 02139}
\date{\today}
\maketitle
\begin{abstract}
We consider the interaction between two rods embedded in a
fluctuating surface. The modification of fluctuations by the rods
leads to an attractive long-range interaction between them.
We consider fluctuations governed by either surface tension
(films) or bending rigidity (membranes). In both cases the
interaction
falls off with the separation of the rods as $1/R^4$. The
orientational
part of the interaction is proportional to
$\cos^2\left[ \theta_1+\theta_2 \right]$ in the former case, and to
$\cos^2\left[ 2\left(\theta_1+\theta_2\right) \right]$ in the latter,
where $\theta_1$ and $\theta_2$ are angles between the rods
and the line joining them. These interactions are somewhat
reminiscent of dipolar forces and will tend to align collections of
such rods into chains.
\end{abstract}
\pacs{87.20, 82.65D, 34.20}
\section{Introduction and Summary}\label{sIntro}
In addition to their structural role of forming the exterior frames
of the cell and its
interior organelles and vesicles, lipid bilayers act as the host and
regulator
of many biophysical and biochemical reactions\cite{alb,gen}.
Inter- and intra-cellular recognition and transport,  adhesion,
regulation of ion
concentrations, and energy conversion, are but a few of the processes
taking
place at the membrane. These tasks are carried out by a variety of
proteins,
glycolipids, and other macromolecules that move through the many
different
lipids that make up the bilayer. The resulting membrane is thus far
from uniform;
there are even examples in which inhomogeneities occur on a larger
scale,
e.g. domains of phase separated lipids or two-dimensional protein
assemblies.
In modeling the physical properties of the cell, it is thus
essential to have a
good understanding of the interactions between inclusions in fluid
membranes.

The pursuit of  ``biologically inspired'' materials, which do not
posses the full
complexity of their natural counterparts, yet retain some of their
useful features,
is quite active. Artificial protein assemblies within lipid membranes
are now routinely produced in the laboratory\cite{Las,CM,Fen}.
Such model-membrane systems have potential applications
for targeted drug delivery and
may also lead to novel applications such as nano-scale pumps,
templates,
functionalized interfaces, and chemical reactors.
The appropriate design of such artificial membranes again requires an
understanding of how inclusions modify the physical properties of the
bilayer,
and how the membrane in turn contributes to the interactions between
inclusions.
The forces between the inclusions can be broadly subdivided into two
categories\cite{isr}. The first category includes interactions that
are present
in the bulk of the solvent. They include the van der Waals
interaction, which
falls off  with separation $R$ as $1/R^6$ at long distances.
The Coulomb interaction is strongly screened under physiological
conditions. (Typical ion concentrations are a few hundred millimolar,
which give a
screening length of less than 10 \AA.)
Hydration and structural forces are also short-ranged.
The second category includes interactions which are mediated by the
membrane
itself: the inclusion disturbs the lipid bilayer and this disturbance
propagates
to neighboring inclusions (c.f. \cite{gen,mou,goul,dan} and
references therein).
When macroscopic thermal fluctuations are unimportant (we refer to
this case
as $T=0$), the resulting interactions tend to be short-ranged.
For example, if in the region around an inclusion the membrane is
forced to deviate from its preferred thickness ($\sim 40$\AA),
the resulting disturbance in the bilayer decays (heals) over
a length of order this thickness \cite{dan}.
Two nearby inclusions then feel an interaction that falls off
exponentially with this characteristic length.

There are also long-range interactions that are mediated by the
membrane.
To describe such interactions, it should be possible to neglect the
microscopic
properties of the membrane, and its molecular lipid bilayer
structure, and focus
on its macroscopic properties. In the long-distance limit,
the membrane is well-described by the elastic Hamiltonian\cite{C,H},
\begin{equation}
{\cal H}=\int \d S \left[\sigma+\frac{\kappa}{2} H^2
+\bar\kappa K \right],\label{CHH}
\end{equation}
where $\d S$ is the surface area element, and $H$, $K$ are the mean
and Gaussian curvatures respectively. The elastic properties of the
surface are then described by the tension $\sigma$, and the bending
rigidities
$\kappa$ and $\bar\kappa$. A finite surface tension is in general the
strongest coupling in Eq.(\ref{CHH}) and dominates the bending
terms at long wavelengths.
This term is present for films on a frame, interfaces at short
distances, and possibly membranes subject to osmotic pressure
differences between their interior and exterior. On the other hand,
for unstressed vesicles, the surface tension is quite small, and may
be neglected at wavelengths well below the size of the vesicle
\cite{DGT,BL,DL,S}. In this case, the energy
cost of fluctuations is controlled by the rigidity terms. For
simplicity we shall refer to surface tension dominated surfaces as
films, and to rigidity controlled ones as membranes.

The long-range interactions between inclusions in a membrane
that result from Eq.(\ref{CHH}) were examined in Ref.\cite{goul}.
If the inclusions are asymmetric across the bilayer and impose a
local curvature, even at $T=0$, there is a long-ranged repulsive
interaction that falls off with distance as $1/R^4$. The energy scale
of this interaction is set by $\kappa$ and $\bar\kappa$.
On the other hand, if thermal fluctuations of the membrane are
included ($T\ne0$), there is a $1/R^4$ interaction for {\it generic}
inclusions. The only requirement is that the rigidity of the
inclusion differs
from that of the ambient membrane\cite{goul}. In particular,
the interaction is attractive if the inclusions are stiffer than the
membrane.
The magnitude of this fluctuation--induced interaction is set by
$k_BT$,
and is totally independent of the rigidities $\kappa$ and
$\bar\kappa$
(\cite{goul}, see also Appendix \ref{Asph}).

\begin{figure}
\epsfysize=2.0truein
\vbox{\vskip 0.15truein \hskip 0.4truein
\epsffile{\dir/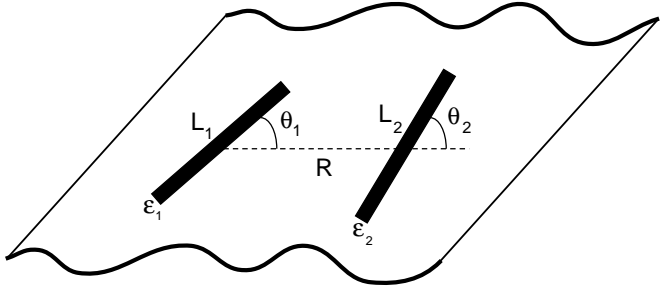}
\vskip 0.15truein
\caption{
Two rod-shaped inclusions embedded in a membrane. The rods are separated
by a distance $R$. The $i$th rod has length $L_i$, width $\epsilon_i$, and
makes an angle $\theta_i$ with the line joining the centers of the two rods.}
\label{fig1}}
\end{figure}

In a recent report, we  considered the dependence of the
fluctuation--induced
 ($T\ne 0$) interaction  between rod-like inclusions on their
orientations\cite{EPL}.
The rods are assumed to be sufficiently rigid so that they do not
deform
coherently with the underlying membrane. They can thus only perform
rigid translations and rotations while remaining attached to the
surface.
As a result, the fluctuations of the membrane are constrained,
having to vanish at the boundaries of the rods.
Consider the situation depicted in Fig.\ \ref{fig1}, with  two rods
of lengths $L_1$ and $L_2$ at a separation $R\gg L_i$. For
fluctuating
films ($\sigma\ne0$), there is an attractive fluctuation-induced
interaction given by,
\begin{equation}\label{film}
V_F^{T}(R,\theta_1,\theta_2)=-\frac{k_{B}T}{128}
\frac{L_1^2 L_2^2}{R^4}\cos^2\left[\theta_{1}+\theta_{2}\right]
+O\left(1/R^6\right),
\end{equation}
where $\theta_1$ and $\theta_2$ are the angles between the
rods and the line adjoining their centers, as indicated in
Fig.\ \ref{fig1}.
This angular dependence is actually the {\it square} of that of
a dipole-dipole interaction in two dimensions, with $L_1$ and $L_2$
as the dipole strengths.
The fluctuation-induced interaction on a membrane  ($\sigma=0$)
is very similar and given by
\begin{equation}\label{memb}
V_M^{T}(R,\theta_1,\theta_2)=-\frac{k_{B}T}{128}
\frac{L_1^2 L_2^2}{R^4}\cos^2\left[2\left(\theta_{1}+\theta_{2}
\right)\right]+O\left(1/R^6\right).
\end{equation}
The orientational dependence is the {\it square}  of a
quadrupole--quadrupole
interaction, with the unusual property of being minimized
for both parallel and perpendicular orientations of the rods. Note
that the
strength of the interaction is the same in both cases.
The above fluctuation-induced interactions decay less rapidly at
large distances than van der Waals forces and may play an
important role in aligning asymmetric inclusions in biomembranes.
Since orientational
correlations are often easier to measure than forces, this result may
also be useful as a probe of the fluctuation-induced interaction.
Finally, this interaction could give rise to novel two-dimensional
structures for collections of rodlike molecules. In particular, the
resemblance of the orientational part of the interaction to dipolar
forces suggests that
a suitable way to minimize the energy of a collection of rods is to
form
them into chains. (If the rods are not colinear, the interactions
cannot be
minimized simultaneously.) Such chain-like structures are observed
for ferromagnetic particles controlled by similar forces\cite{ros}.

In this article we provide the calculational details that lead to the
above results.
The remainder of the paper is organized as follows. Due to its richer
complexity,
the detailed calculation for the force between rods on membranes is
presented first in Sec.\ref{smemb}. The corresponding calculation for
films
is described more succinctly in Sec.\ref{sfilm}. This is followed by
a discussion of the results, and comparison with other work in
Sec.\ref{sdisc}. Various calculational details are relegated to the
appendices. Asymmetric inclusions have also been recently considered
in \cite{PL}, where results similar to ours are reported.

\section{Membranes}\label{smemb}
We start with a thermally fluctuating planar membrane subject to
the Hamiltonian in Eq.(\ref{CHH}). We assume that the size of the
membrane
$d$, is well below the persistence length $\xi$\cite{DGT}. In this
limit,
the membrane undergoes only small fluctuations about a flat state.
We may then parametrize the membrane surface with a height function
$u(r)$ and approximate the full coordinate-invariant Hamiltonian of
Eq.(\ref{CHH}) by the quadratic form
\begin{equation}
{\cal H}_0=\frac{\kappa}{2} \int_{{\rm IR}^2}\d^2r
\left(\nabla^{2}u(r)\right)^2.\label{ham}
\end{equation}
Since we assume $d$ is large (compared with $R$ and $L_i$), we denote
the (finite but large) reference plane by ${\rm IR}^2$.

Now consider the situation depicted in Fig.\ \ref{fig1}, where two
rigid, rod-like objects, are attached to the membrane.
We shall represent the rods by narrow rectangles
of lengths $L_1$ and $L_2$, and widths $\epsilon_1$ and
$\epsilon_2$; ultimately taking the limit of $\epsilon_i\to0$.
The rods are constrained to
fluctuate with the membrane but, due to their stiffness, can only
be tilted or translated up and down rigidly. We can parametrize
all possible configurations of the rods by
\begin{eqnarray}
{\hskip 4cm} u(r)|_{r\in L_i}=a_i+{\bf{b}}_i\cdot{\bf{r}},
{\hskip 1.5cm}{\rm for}\quad i=1,2\qquad,  \label{abdef}
\end{eqnarray}
where we have also used $L_i$ to denote the $i$th rod.
To calculate the partition function, we follow a procedure similar
to Ref.\cite{goul} and sum over all possible configurations of the
membrane,
weighted by the corresponding Boltzmann factor, and subject to the
constraints imposed by the rods via Eq.(\ref{abdef}). The constraints
may be implemented with the aid of delta functions as
in Ref.\cite{LiK}, leading to
\begin{equation}
{\cal Z}=\int {\cal D}u(r) \prod_{i=1}^{2} \int \d a_i
\d^2b_i
	\prod_{r'\in{L_i}}
\delta\left(u(r')-a_i-{\bf{b}}_i\cdot {\bf{r}}'\right)
	\exp\left[-\frac{{\cal H}_0}{k_{B}T}\right]. \label{Z1}
\end{equation}
In Eq.(\ref{Z1}) we have included only the leading term in an
expansion
in powers of ${\bf b}_i$. As described in Appendix \ref{Atilt},
higher order terms
come from the projection of $L_i$ onto the $x-y$ plane, as well as
from the
integration measure for ${\bf b}_i$, which is on the sphere of unit
normals.
Since ${\bf b}_i$ controls the gradient of $u(r)$ at the boundary of
$L_i$,
the expansion in ${\bf b}_i$ is in the same spirit as the gradient
expansion
for the Hamiltonian in Eq.(\ref{ham}). In Appendix \ref{Atilt} we
further demonstrate
that, just as in the case of  anharmonic terms that have been
neglected in
Eq.(\ref{ham}), the higher order terms in ${\bf b}_i$ left out from
Eq.(\ref{Z1})
are suppressed in the  limit $d\ll\xi$.
Expressing the delta functions as functional integrals over auxiliary
fields $k_i(r)$ defined on the rods, we obtain
\begin{eqnarray}
\lefteqn{{\cal Z}=\int {\cal D}u(r) \prod_{i=1}^2 \int \d a_i
\d^2b_i\int {\cal D}k_i(r) }\hskip.5cm\label{Z2}\\
 & & \times\;\exp\left[-\frac{\kappa}{2k_{B}T}
\int_{{\rm IR}^2} \d^2r \left(\nabla^{2}u(r)\right)^2+i
\sum_{i=1}^2\int_{L_i}
\d^2r_i k_i(r_i)
\left(u(r_i)-a_i-{\bf{b}}_i\cdot{\bf{r}}_i\right)\right]
 .\nonumber
\end{eqnarray}
Integrating out $u(r)$,  $a_i$, and ${\bf b}_i$, then gives
\begin{eqnarray}
\lefteqn{{\cal Z}=\prod_{i}\int{\cal D}k_i(r)\delta\left(\int_{L_i}
\d^2r_ik_i(r_i)\right)\delta^2\left(\int_{L_i}\d^2r_i\,{\bf
r}\,k_i(r_i)
\right)}\hskip.5cm  \nonumber\\
 & &
\times\;\exp\left[-\frac{k_BT}{2\kappa}\sum_{i,j=1}^2\int_{L_i}\d^2r_i
\int_{L_j}\d^2r_j
k_i( r_i)G({\bf r}_i-{\bf r}_j)k_j(r_j)\right],\label{coul}
\end{eqnarray}
where
\begin{equation}
 G({\bf r}-{\bf r}')=\left(\frac{1}{\nabla^4}\right)_{{\bf rr}'}
=\frac{1}{8\pi}\mid{\bf r}-{\bf r}'\mid^2\ln\mid{\bf r}-{\bf
r}'\mid.\label{gf}
\end{equation}
Equation (\ref{coul}) is analogous to the partition function for a
pair of
plasmas confined to the interior of rods $L_1$ and $L_2$. The delta
functions impose the constraints that the net charge and dipole
moments vanish within each rod. When the distance $R$ between rods
is much bigger than their size (i.e. $L_i\ll R$), we may approximate
$G({\bf r}_1-{\bf r}_2)$ in  Eq.(\ref{coul}) by a multipole expansion
and keep only the leading term, which comes from
the quadrupole moments (see Appendix \ref{AQQ})
\begin{equation}
Q^{(i)}_{ab}\equiv\int_{L_i}\d^2r\;r_a r_b\, k_i(r).
\end{equation}
After inserting
\begin{equation}
 1=\prod_{i=1}^2\int \d{\bf Q}^{(i)}\d{\bf
g}^{(i)}\;\exp\left[i\sum_{ab}g^{(i)}_{ab}
\left(Q^{(i)}_{ab}-\int_{L_i}\d^2r\; r_a r_b \,
k_i(r)\right)\right],\label{gdef}
\end{equation}
into Eq.(\ref{coul}) and performing the multipole expansion, we
obtain
\begin{eqnarray}
{\cal Z}&=&\prod_{i}\int{\cal D}k_i(r)\int \d{\bf Q}^{(i)}d{\bf
g}^{(i)}\d a_i
\d^2b_i \nonumber\\
&&\times\;\exp\left\{-\frac{k_BT}{2\kappa}\sum_{i}\int_{L_i}
\d^2r\d^2r'k_i(r)G({\bf r}-{\bf r'})k_i(r')\right.\nonumber\\
&&-i\sum_{i}\int_{L_i}\d^2r k_i(r)\left[a_i+{\bf b}_i\cdot{\bf r}+
{\bf r}\cdot{\bf g}^{(i)}\cdot{\bf r}\right]\nonumber\\\label{ZK}
&&\left. +i\sum_{i}g^{(i)}_{ab}Q^{(i)}_{ab}-\frac{k_BT}{2\kappa}
v\left[{\bf Q}^{(1)},{\bf Q}^{(2)}\right]\right\},
\end{eqnarray}
where we have recast the delta functions in Eq.(\ref{coul}) in terms
of integrals over $a_i$, and ${\bf b}_i$. The quadrupole-quadrupole
interaction
\begin{eqnarray}
v\left[{\bf Q}^{(1)},{\bf Q}^{(2)}\right]&=&\frac{1}{8\pi
R^2}\left[Q^{(1)}_{aa}Q^{(2)}_{bb}
+2Q^{(1)}_{ab}Q^{(2)}_{ab}-2Q^{(1)}_{aa}\hat{{\bf R}}\cdot{\bf
Q}^{(2)}\cdot\hat{{\bf R}}
-2Q^{(2)}_{aa}\hat{{\bf R}}\cdot{\bf Q}^{(1)}\cdot\hat{{\bf
R}}\right.
\nonumber\\
&&-8\hat{{\bf R}}\cdot{\bf Q}^{(1)}\cdot{\bf Q}^{(2)}\cdot\hat{{\bf
R}}
+8\hat{{\bf R}}\cdot{\bf Q}^{(1)}\cdot\hat{{\bf R}}\;
\left.\hat{{\bf R}}\cdot{\bf Q}^{(2)}\cdot\hat{{\bf R}}\right]+{\rm
O}(1/R^3),  \label{vqq}
\end{eqnarray}
(with implicit summation over repeated $a$ and $b$)
is calculated in Appendix \ref{AQQ}.
Note that the Green's function in Eq.(\ref{gf}) should also contain
homogeneous terms, which reflect the boundary conditions
at the outer edge of the membrane, $r=d$. However, we have only
used the explicit form of the Green's function in computing the
leading terms in the multipole expansion. As long as $L_1$ and
$L_2$ are sufficiently far (compared to $R$) from the edge,
the particular choice of boundary conditions at  $r=d$ does not
modify the leading terms in this expansion. The homogeneous terms
can therefore be safely suppressed in Eq.(\ref{gf}).

We first isolate the integration over $k_1(r)$ in Eq.(\ref{ZK}),
\begin{eqnarray}
I_1&\equiv&\int{\cal
D}k_1(r)\exp\left\{-\frac{k_BT}{2\kappa}\int_{L_1}\d^2r
\d^2r'
k_1(r)G({\bf r}-{\bf r}')k_1(r')\right.\nonumber\\
&&\left. -i\int_{L_1}\d^2r k_1(r) \left[a_1+{\bf b}_1\cdot{\bf r}+
{\bf r}\cdot{\bf g}^{(1)}\cdot{\bf r}\right]\right\}.\label{Idef}
\end{eqnarray}
To perform the above integration, the Green's function in
Eq.(\ref{gf})
has to be inverted in the finite region $L_1$. In order to do this,
we introduce an auxiliary field $h({\bf r})$ and write
\begin{eqnarray}
I_1&=&\int{\cal D}h(r)\exp\left[-\frac{\kappa}{2k_BT}\int_{{\rm
IR}^2}
\d^2r
\left(\nabla^2 h(r)\right)^2\right]\nonumber\\
&&\times\prod_{r'\in L_1}\delta\left(h(r')-a_1-{\bf b}_1\cdot{\bf r}-
{\bf r}\cdot{\bf g}^{(1)}\cdot{\bf r}\right).
\end{eqnarray}
This is just the partition of a membrane constrained by a single
curved rod. After evaluating the contribution on $L_1$
(via the delta function), we are left with
\begin{equation}
I_1=\exp\left[-\frac{2\kappa}{k_BT}\,\epsilon_1L_1\,\left(g^{(1)}_{aa}
\right
)^2
\right]
\int'{\cal D}h(r) \exp\left[-\frac{\kappa}{2k_BT}\int_{{\rm
IR}^2-L_1}
\d^2r\, \left(\nabla^2 h(r)\right)^2\right],
\end{equation}
where the prime indicates that the function $h(r)$, and its normal
gradient, are  constrained to satisfy the following conditions on
the boundary $\partial L_1$, of $L_1$,
\begin{eqnarray}
\left. h(r)\right|_{r\in\partial L_1}=a_1+{\bf b}_1\cdot{\bf r}+
{\bf r}\cdot{\bf g}^{(1)}\cdot{\bf r},
\nonumber\\
\left. \frac{\partial h(r)}{\partial {\bf n}}\right|_{r\in\partial
L_1}
=\frac{\partial}{\partial {\bf n}}\left(a_1+{\bf b}_1\cdot{\bf r}+
{\bf r}\cdot{\bf g}^{(1)}\cdot{\bf r}\right).\label{bc}
\end{eqnarray}
Now let $h_0(r)$ denote a solution to the biharmonic equation,
\begin{equation}
\nabla^4 h_0=0 \label{biharm}
\end{equation}
on ${\rm IR}^2-L_1$
with the boundary conditions of Eqs.(\ref{bc}). We then set
\[ h(r)=h_0(r)+\delta h(r),\]
where both $\delta h(r)$, and its normal gradient, are zero on the
boundary
of $L_1$. Following this change of variables,
\begin{equation}
I_1=A
 \exp\left[-\frac{2\kappa}{k_BT}\,\epsilon_1L_1\,\left(g^{(1)}_{aa}
 \right)^2
 \right]
\exp\left[-\frac{\kappa}{2k_BT}\int_{{\rm IR}^2-L_1}
\d^2r\, \left(\nabla^2 h_0(r)\right)^2\right],\label{I1}
\end{equation}
where
\[
A=\int{\cal D}\delta h(r)\exp\left[-\frac{\kappa}{2k_BT}\int_{{\rm
IR}^2-L_1}
\d^2r \left(\nabla^2 \delta h(r)\right)^2\right],
\]
is a normalization constant, independent of $a$, ${\bf b}$,
and  ${\bf g}$, which does not affect the remaining computations.
In order to solve Eq.(\ref{biharm}) we must specify the boundary
conditions at $r=d$, which are the same as those for $u(r)$.
As discussed earlier, the results should be independent of this
choice,
and it is convenient to select
\begin{equation}
h|_{r=d}= {\partial h\over\partial r}\Big|_{r=d}=0.
\end{equation}
As shown in Appendix \ref{Abih}, the solution for the case when the
rod is along the $y$-axis, in the limit $d\gg L_1$, gives
\begin{eqnarray}
\lefteqn{\int_{{\rm IR}^2-L_1}  d^2r \left(\nabla^2
h_0(r)\right)^2}\label{S}\\
    & &=2\pi\left(L_1g^{(1)}_{xy}\right)^2+\frac{1}
{\ln(4d/L_1)}\left[s_1
b^2_{1x} +s_2 b^2_{1y} \right] +O(L_1/d),\nonumber
\end{eqnarray}
where $s_i$ are numerical constants. The second term on the
right hand side of Eq.(\ref{S}) is examined in Appendix  \ref{Atilt},
where the irrelevance of higher order terms in ${\bf b}$
is demonstrated. In the limit $d\gg L_1$, it suffices to keep only
the
first term on the right-hand side of Eq.(\ref{S}), which gives
\begin{equation}
I_1=A\exp\left\{-\frac{\kappa}{k_BT}\left[2\epsilon_1L_1\,
\left(g^{(1)}_{aa}\right)^2 +\pi\left(L_1
g^{(1)}_{xy}\right)^2\right]\right\}.
\label{I}
\end{equation}
The result of the $k_2(r)$ integration in Eq.(\ref{ZK}) is similar,
with the index 2 replacing 1, and with the coordinate axis
appropriately rotated to align with the second rod.
The overall expression for the partition function now reads
(dropping unimportant multiplicative constants)
\begin{eqnarray}
{\cal Z}&=&\prod_{i=1}^2 \int \d{\bf Q}^{(i)} \d{\bf g}^{(i)}
\exp\left\{-\frac{\pi\kappa}{k_BT}\left[\left(L_1\;g^{(1)}_{x'y'}
\right)^2
+\left(L_2\;g^{(2)}_{x''y''}\right)^2\right]\right\} \\
& &\hskip 2cm \times\;\exp\left\{-i\sum_{i} g^{(i)}_{ab} Q^{(i)}_{ab}
-\frac{k_BT}{2\kappa}
v\left[{\bf Q}^{(1)},{\bf Q}^{(2)}\right]\right\} \nonumber,
\end{eqnarray}
where we have set the widths of the rods to zero (i.e. taken the
$\epsilon_i \to0$ limit). The primed indices $x',y',x'',y''$ indicate
that the corresponding components are with respect to the
coordinate frames where $L_1\parallel y'$ and $L_2\parallel y''$.
We define an un-primed coordinate system such that the $x$-axis
is parallel to $\hat{R}$ and the two rods make angles of $\theta_1$
and $\theta_2$ with respect to the $x$-axis as in Fig.\ \ref{fig1}.
Integration over ${\bf g}$ yields
\begin{eqnarray}\label{lastz}
{\cal Z}&=&\prod_{i=1}^2 \int \d{\bf Q}^{(i)}\;
\delta\left(Q^{(i)}_{xx} \cos^2\theta_i +Q^{(i)}_{xy} \sin2\theta_i
+Q^{(i)}_{yy} \sin^2\theta_i\right) \\
& &\hskip 1.5cm \times\;
\delta\left(Q^{(i)}_{xx} \sin^2\theta_i -Q^{(i)}_{xy} \sin2\theta_i
+Q^{(i)}_{yy} \cos^2\theta_i\right) \nonumber \\
& & \times\;
\exp\left\{-\frac{k_BT}{2\kappa}\left[\sum_{i}\frac{1}{2\pi L_i^2}
\left(\frac{1}{2}(Q^{(i)}_{yy}-Q^{(i)}_{xx})\sin2\theta_i
+Q^{(i)}_{xy}\cos2\theta_i\right)^2\right]\right\}   \nonumber\\
& &\hskip 2cm \times\;\exp\left\{-\frac{k_BT}{2\kappa}
v\left[{\bf Q}^{(1)},{\bf Q}^{(2)}\right]\right\} \nonumber.
\end{eqnarray}

Since we are working in the large-$R$ limit, the
${\bf Q}$ integrations are most easily performed by expanding
Eq.(\ref{lastz}) to second order in $v$. After expanding
$-k_BT\log {\cal Z}$, we find the $(R,\theta_1,\theta_2)$-dependent
part of the free energy given in Eq.(\ref{memb}).
We can re-write this interaction in a coordinate invariant form,
in terms of the vector $\hat{R}$ and the directors $\hat{L}_{1}$
and $\hat{L}_{2}$ along the rods as
\begin{equation}
V_M^{T}=-\frac{k_{B}T}{128}{L_1^2L_2^2\over R^4}
\left[2\left(2 (\hat{L}_{1} \cdot \hat{R})(\hat{L}_{2} \cdot \hat{R})
-\hat{L}_{1} \cdot \hat{L}_{2} \right)^2 -1 \right]^2
+O(1/R^6).
\end{equation}

\section{Films}\label{sfilm}
We now turn to the analogous calculation for films. Again we use a
Gaussian approximation for the Hamiltonian in Eq. (\ref{CHH}),
but keep only the surface tension this time,
\begin{equation}
{\cal H}_0=\frac{\sigma}{2} \int\d^2r
\left(\nabla u(r)\right)^2.\label{fham}
\end{equation}
All anharmonic corrections to the above expression are unimportant
in the limit $\sigma a^2 \gg k_{B}T$, where $a$ is a microscopic
length. We follow a procedure similar to that described in
Sec.(\ref{smemb}) but with the differences noted below.
The expression for the partition function is similar to
Eq.(\ref{Z1}),
with ${\cal H}_0$ now given by Eq.(\ref{fham}). For films however, we
cannot justify keeping only the leading terms in an expansion in ${\bf
b}_i$. Thus the full rotationally-invariant measure of integration on
the sphere of slopes
${\bf b}_i$ should be employed (see Appendix \ref{Atilt}). Also, the
appropriate domain replacing $L_i$ is the projected length $\bar
{L_i}\equiv
L_i/\sqrt{1+b_{iy}^2}$. After introducing auxiliary fields $k_i(r)$
as
in Sec.(\ref{smemb}), the analog of Eq.(\ref{coul}) may be written as
\begin{eqnarray}
\lefteqn{{\cal Z}=\prod_{i}
\int \frac{\d^2 b_i}{\left(1+{\bf b}_i^2\right)^{3/2}}
\int{\cal D}k_i(r)
\delta\left(\int_{\bar {L_i}} \d^2r_ik_i(r_i)\right) }
\hskip.5cm  \nonumber\\
 & & \times\;\exp\left[-\frac{k_BT}{2\sigma}\sum_{i,j=1}^2\int_{\bar
{L_i}}
\d^2r_i\int_{\bar L_j}\d^2r_j
k_i( r_i){\cal G}({\bf r}_i-{\bf r}_j)k_j(r_j)
-i \sum_i {\bf b}_i \cdot \int_{\bar {L_i}} \d^2r_i {\bf r}_i
k_i(r_i) \right]
,\label{fcoul}
\end{eqnarray}
where
\begin{equation}
 {\cal G}({\bf r}-{\bf r}')=\left(\frac{1}{-\nabla^2}\right)_{{\bf
rr}'}
=\frac{1}{2\pi} \ln\mid{\bf r}-{\bf r}'\mid.\label{fgf}
\end{equation}
Note that for films, the dipole moment of $k_i(r)$ does not vanish.
Expanding ${\cal G}({\bf r}-{\bf r}')$ in a multipole expansion and
keeping
only the leading term, which now comes from the dipole moments
${\bf p}_i\equiv\int{\bf r}k_i(r)$, we find
\begin{eqnarray}
{\cal Z}&=&\prod_{i}\int{\cal D}k_i(r)\int \d{\bf p}^{(i)}d{\bf
g}_i\d a_i
\frac{\d^2 b_i}{\left(1+{\bf b}_i^2\right)^{3/2}} \nonumber\\
&&\times\;\exp\left\{-\frac{k_BT}{2\sigma}\sum_{i}
\int_{\bar {L_i}}\d^2r\d^2r'k_i(r)
{\cal G}({\bf r}-{\bf r'})k_i(r')\right.\nonumber\\
&&-i\sum_{i}\int_{\bar {L_i}}\d^2r k_i(r)\left[a_i+({\bf b}_i+{\bf
g}_i) \cdot
{\bf r}\right]\nonumber\\\label{fZK}
&&\left. +i\sum_{i}{\bf g}_i \cdot {\bf p}^{(i)}-\frac{k_BT}{2\sigma}
u\left[{\bf p}^{(1)},{\bf p}^{(2)}\right]\right\},
\end{eqnarray}
where
\begin{equation}
u\left[{\bf p}^{(1)},{\bf p}^{(2)}\right]=-\frac{1}{\pi R^2}
\left[{\bf p}^{(1)} \cdot {\bf p}^{(2)}
-2 ({\bf p}^{(1)} \cdot \hat{R}) ({\bf p}^{(2)} \cdot \hat{R})
\right],
\label{vpp}
\end{equation}
is the effective dipole-dipole interaction, analogous to
Eq.(\ref{vqq}),
and ${\bf g}_i$ is the variable conjugate to ${\bf p}^{(i)}$.
We integrate out $k_i(r)$ as in Sec.(\ref{smemb}) by introducing an
auxiliary field $h(r)$. In this case we must solve a harmonic
equation on ${\rm IR}^2-\bar {L_i}$, instead of the biharmonic
Eq.(\ref{biharm}), with the
boundary condition on $\bar {L_i}$
\begin{equation}
\left. h_0(r)\right|_{r\in\partial \bar {L_i}}=a_i+({\bf b}_i+{\bf
g}_i) \cdot
{\bf r}\quad. \label{fbc}
\end{equation}
The harmonic problem can be solved either by a method similar to that
described in Appendix \ref{Abih}, or by conformal mapping.
The resulting expression for the partition function reads
\begin{eqnarray}
{\cal Z}= \prod_i \int &&\d {\bf p}^{(i)} \d {\bf g}_i
\frac{\d^2 b_i}{\left(1+{\bf b}_i^2\right)^{3/2}}     \nonumber \\
&&\times\;\exp\left[-\frac{\sigma}{2k_{B}T}
\left(\frac{\pi}{4}\bar {L_1}^2\left(b_{1y'}+g_{1y'}\right)^2
+\frac{\pi}{4}\bar {L_2}^2 \left(b_{2y''}+g_{2y''}\right)^2 \right)
+i \sum_i {\bf g}_i \cdot {\bf p}^{(i)}-\frac{k_BT}{2\sigma}
u\left[{\bf p}^{(1)},{\bf p}^{(2)}\right]\right],\label{fZ2}
\end{eqnarray}
where the meaning of the primes is the same as in
the previous section.
One can now see explicitly that the higher order terms in the
expansion
in ${\bf b}_i$ are important in this case. The remaining
integrations,
except those of  $b_{1y'}$ and $b_{2y''}$, can be performed in
a straightforward manner. The latter two integrals are rather
complicated and in
order to get a simple result, we restrict to the case $\sigma L^2
\left({L^2}/{R^2}\right) \ll k_{B}T $. In this limit, the integrals
can be approximated by Gaussian forms. After expanding $-k_BT\log
{\cal Z}$, we find Eq.(\ref{film}), which can  be expressed in a
coordinate invariant
form, in terms of the vector $\hat{R}$ and the directors
$\hat{L}_{1}$
and $\hat{L}_{2}$ along the rods, as
\begin{equation}
V_F^{T}=-\frac{k_{B}T}{128} \frac{L_1^2 L_2^2}{R^4}
\left[2 (\hat{L}_{1} \cdot \hat{R})(\hat{L}_{2} \cdot \hat{R})
-\hat{L}_{1} \cdot \hat{L}_{2} \right]^2
+O(1/R^6).
\end{equation}

\section{Discussion}\label{sdisc}
We shall now discuss some general aspects of the fluctuation-induced
interactions in Eqs.(\ref{film}) and (\ref{memb}). The
magnitude of the interaction is solely determined by $k_BT$ and is
independent of the tension and rigidity coefficients $\sigma$ and
$\kappa$. Thus the effect persists even for rather stiff membranes
with $\kappa\gg k_BT$. The only assumption is that the inclusions
are much more rigid than the embedding surface, thus limiting its
fluctuations in their neighborhood.

For both membranes and films, the interaction falls off with distance
as $1/R^4$. This is a general feature of fluctuation-induced forces,
including the (finite $T$) van der Waals interactions, which in $d$
dimensions fall off as $1/R^{2d}$. Since the direct van der Waals
interactions between inclusions still fall off as $1/R^6$, the forces
mediated through the two-dimensional surface will always
asymptotically dominate. Of course the dimensional dependence of
$R^4$ is canceled by a product of lengths in the numerator.
For spherical inclusions, this is given by the product of two
inclusion areas (see Ref.\cite{goul} and Appendix \ref{Asph}) and
for rods by the product of the squares of their lengths.
Presumably, for general shapes, there
is a formula that interpolates between these two limits.
Another potential extension is to a polymer floating on a membrane.
The interplay between the elasticity and shapes of a polymer and
membrane, neglecting membrane fluctuations, have been examined in
\cite{Kozlov}; an extension to the case of a fluctuating membrane has
also recently appeared\cite{Podgornik}. There is also interesting
behavior in the opposite limit of $R\ll L$ for the interaction between
two parallel semiflexible polymers\cite{ramin}.

\begin{figure}
\epsfysize=2.0truein
\vbox{\vskip 0.15truein \hskip 0.4truein
\epsffile{\dir/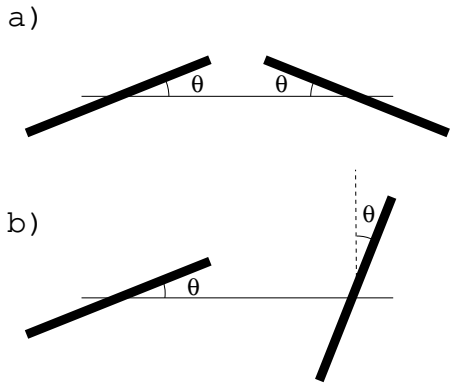}
\vskip 0.15truein
\caption{The minimal-energy orientations for two rods in a membrane (a and b)
and a film (a only). The energy is minimal for all values of $\theta$.}
\label{fig2}}
\end{figure}

Finally, the most interesting aspect of our calculation is the
orientational dependence of the force. This is most easily discussed
for the film, where an intermediate stage involves calculating the
angular
dependence of a dipole-dipole interaction, which is subsequently
squared. The final angular dependence is thus that of {\it squared
dipolar interactions}. Similarly, the result for the membrane
corresponds to {\it squared quadrupolar interactions}.
The minimal-energy orientations
are shown in Fig.\ \ref{fig2}; note there is a large degeneracy.
We also note that these interactions cannot be obtained by adding
two-body potentials on the rods: To find the orientational
dependence
of additive forces, let us consider an interaction
$U(|r_1-r_2|)du_1du_2$,
between any two infinitesimal segments of two rods of length $L$ at a
distance $R\gg L$. Expanding $ |r_1-r_2|$, and integrating over
the two rods, leads to the interaction
\begin{equation}
V(R,\theta_1,\theta_2)=L^2U(R)+{L^4 \over 6}\left( {U'(R) \over R}
+U''(R)\right)
-\frac{L^4}{12}\left(  {U'(R) \over R}
-U''(R)\right)(\cos2\theta_{1}+\cos2\theta_{2}).\label{add}
\end{equation}
Note that the angular dependence is now completely different, and
minimized when the two rods are parallel to their axis of separation.
Presumably both interactions are present for rods of finite thickness;
the additive interaction is proportional to $L^2(L\epsilon/R)^2$,
where
$\epsilon$ is the thickness. The previously calculated interactions
are thus larger by a factor proportional to $(R/\epsilon)^2$ and
should dominate at large separations.

The unusual dependence on orientation in Eqs.(\ref{film}) and
(\ref{memb}) could lead to new types of orientational ordering in
ensembles of rod-shaped particles. Of course, due to the non-additive
nature of the forces, the fluctuation-induced interaction should
be calculated separately for each arrangement. However, a cursory
examination suggests that three-body and higher order interactions
fall off with separation as $1/R^6$. Thus for $R\gg L$, a collection
of rods can be treated as if they interact through additive pair
potentials.
It is amusing to examine the minimum of such an interaction for three
rods placed on the vertices of an equilateral triangle. One possible
equilibrium configuration is a three arm star with the relative angles
of $2\pi/3$ between the rods.  Interestingly, this so-called
``triskelion'' structure is indeed formed by three rod-like
``clathrin'' proteins\cite{alb}.
(Another stable configuration has each rod parallel
to the corresponding side of the equilateral triangle.)
Of course, given the relative proximity of the three proteins,
it is not clearthat the asymptotic interactions of Eq.(\ref{memb}) are
applicable to this
case.  Another generic aspect of dipole and quadrupolar interactions
is that they are frustrated (i.e. cannot be simultaneously minimized
with respect to the orientations) if the rod centers are not aligned.
There may thus be an overall tendency to arrange rod shaped
molecules into chains. (Naturally this effect competes with the
tendency to aggregate the inclusions together.)
We hope that the orientational dependent interactions calculated
in this paper will provide a fresh perspective on the behavior of
inclusions in biological membranes.

\acknowledgements
MK and MG acknowledge the hospitality of the ITP at Santa Barbara
where this work was initiated (supported by NSF Grant No.
PHY-89-04035).
The work at MIT is supported by the NSF grant DMR-93-03667.
RG acknowledges support from the Institute for Advanced Studies in
Basic
Sciences, Gava Zang, Zanjan, Iran.

\appendix
\section{The quadrupole-quadrupole interaction}\label{AQQ}

In this appendix we derive the quadrupole-quadrupole
interaction in Eq.(\ref{vqq}). The starting point is the pairwise
interaction
\begin{equation}\label{a1}
v=2 \int_{L_1}\d^2r \int_{L_2}\d^2r'\, k_1(r) G(r-r'-R) k_2(r').
\end{equation}
The Green's function may be written as
\begin{eqnarray}
G({\bf r}-{\bf r}'-{\bf R})&=&\frac{1}{8\pi}\mid{\bf r}-{\bf r}'
-{\bf R} \mid^2 \ln\mid{\bf r}-{\bf r}'-{\bf R} \mid \\
&=&\frac{1}{16\pi} R^2 \left(1+\frac{r^2}{R^2}+\frac{r'^2}{R^2}
-2\frac{{\bf r} \cdot {\bf r'}}{R^2}-2\frac{{\bf r} \cdot \hat{\bf
R}}{R}
+2\frac{{\bf r'} \cdot \hat{\bf R}}{R} \right) \\
& & \times\; \left[ \ln \left(1+\frac{r^2}{R^2}+\frac{r'^2}{R^2}
-2\frac{{\bf r} \cdot {\bf r'}}{R^2}-2\frac{{\bf r} \cdot \hat{\bf
R}}{R}
+2\frac{{\bf r'} \cdot \hat{\bf R}}{R} \right) +\ln(R^2) \right].
\nonumber
\end{eqnarray}
Since ${\bf r}$ and ${\bf r'}$ are limited to the interior of
$L_1$ and $L_2$, we can expand the above expression in powers of the
small quantities $|{\bf r}/R|$, $|{\bf r'}/R|$.
Because of the constraints
\begin{eqnarray}
\int_{L_i} \d^2r k_i(r)=0 ,  \nonumber\\
\int_{L_i} \d^2r\, r_a\, k_i(r)=0   ,
\end{eqnarray}
(for $i=1,2$) the leading terms in the integration vanish in
Eq.(\ref{a1}).
The first non-vanishing term comes from the quadrupole moment
\begin{equation}
\int_{L_i} \d^2r\, r_a r_b\, k_i(r)=Q^{(i)}_{ab},
\end{equation}
and is given by Eq.(\ref{vqq}).

\section{Integration over tilt angles}\label{Atilt}

In this appendix we examine the higher order terms in the tilts of the
rods ${\bf b}_i$, and show that they may be neglected. For simplicity
we shall focus on  ${\bf b}_1$; similar arguments apply to
${\bf b}_2$.
Whenever possible we drop the index and use ${\bf b}\equiv {\bf b}_1$
and $L\equiv L_1$. The integration for ${\bf b}$ must be performed
over all possible orientations of the rod $L$ in the three
dimensional
embedding space. The manifold of orientations is the unit sphere.
In terms of  the vector ${\bf b}$, defined in Eq.(\ref{abdef}),
the rotation invariant measure on the unit sphere is given by
\begin{equation}
\d\Omega=\frac{\d^2 b}{\left(1+{\bf
b}^2\right)^{3/2}}\label{bmeas}\,.
\end{equation}
The leading term of the expansion of Eq.(\ref{bmeas}) in ${\bf
b}$,  $\d^2 b$,
was used as the integration measure in Eq.(\ref{Z1}).
Additional ${\bf b}$ dependence comes from the projection of the
tilted rods
onto the $x-y$ reference plane. For example, the conditions imposed
in Eq.(\ref{abdef}) do not really apply for $r\in L$ but rather for
$r$ in
the projected image of $L$, which is a rod of length
$L/(1+b_y^2)^{1/2}$. Again, in Eq.(\ref{Z1}) we have taken the
leading order in an expansion in ${\bf b}$ by setting
$1/(1+b_y^2)^{1/2}\approx1$.

We shall now  demonstrate that the higher order terms in
${\bf b}$ can be neglected (as discussed after Eq.(\ref{Z1})).
The argument is presented explicitly for terms of order  ${\bf b}^2$,
but is easily extended to higher orders. Including the first
corrections to Eq.(\ref{Z2}) results in
\begin{eqnarray}
\lefteqn{{\cal Z}\equiv\int {\cal D}u(r) \prod_{i=1}^2
{\cal D}k_i(r)\d a_i\d^2 b_i
\;\left(1+\Gamma_xb_x^2+\Gamma_yb_y^2\right)} \label{Z1corr}\\
 & & \times\;\exp\left[-\frac{\kappa}{2k_{B}T}
\int \d^2r \left(\nabla^{2}u(r)\right)^2+i \sum_i\int_{L_i}
\d^2r_i k_i(r_i) \left(u(r_i)-a_i-{\bf
b}_i\cdot{\bf{r}}_i\right)\right]
 ,\nonumber
\end{eqnarray}
where $\Gamma_{x}$ and $\Gamma_{y}$ are independent of
${\bf b}$. If, as in Sec.\ref{smemb}, only the leading term is
retained,
the integration over ${\bf b}$ leads to the constraint that the
dipole
moment $k_1(r)$ must be zero (see Eqs.(\ref{Z2},\ref{coul})).
Due to the higher order terms in ${\bf b}$, this constraint is
modified,
and we have to take into account  the dipole moment
\begin{equation}
{\bf p}\equiv\int_{L_1}\d^2r\,{\bf r}k_1(r).
\end{equation}
Following the same procedure used for the quadrupole moment in
Sec.\ref{smemb}, we introduce an auxiliary variable ${\bf f}$, via
\begin{equation}
 1=\int \d{\bf p}\d{\bf f}\;\exp\left[i\,{\bf f}\cdot
\left({\bf p}-\int_{L_1}\d^2r\; {\bf r}\,
k_1(r)\right)\right].\label{fdef}
\end{equation}
Inserting Eqs.(\ref{fdef}) and (\ref{gdef}) into Eq.(\ref{Z1corr}),
and
performing the multipole expansion, gives

\begin{eqnarray}
{\cal Z}&=&\int\prod_{i}{\cal D}k_i(r)\d{\bf Q}^{(i)}\d a_i
\d{\bf g}^{(i)}\d{\bf b}_2\d{\bf b}\d{\bf f} \d{\bf p}\;
\left(1+\Gamma_xb_x^2+\Gamma_yb_y^2\right)\nonumber\\
&&\times\;\exp\left\{-\frac{k_BT}{2\kappa}\int_{L_1}\d^2r\d^2r'k_1(r)
G({\bf r}-{\bf r'})k_1(r')-i({\bf b}-{\bf f})\cdot{\bf
p}\right.\nonumber\\
&&\left.-i\int_{L_1}\d^2r k_1(r)\left[a_1+{\bf f}\cdot{\bf r}+
{\bf r}\cdot{\bf g}^{(1)}\cdot{\bf r}\right]\right\}\times\cdots\,.
\end{eqnarray}
In the above equation, $\cdots$ denotes factors that are independent
of
$k_1(r)$, ${\bf b}$, and ${\bf f}$, and identical to the
corresponding
terms in Eq.(\ref{ZK}) with the exception that
$v\left[{\bf Q}^{(1)},{\bf Q}^{(2)}\right]$ is replaced by
$v\left[{\bf p},{\bf Q}^{(1)},{\bf Q}^{(2)}\right]$, i.e. the
multipole
energy now also depends on ${\bf p}$.
The integration over $k_1(r)$ is the same as in Eq.(\ref{Idef}),
except
that ${\bf b}_1$ is replaced by ${\bf f}$. Thus, after substituting
${\bf f}$ for ${\bf b}_1$ in Eq.(\ref{S}), we are left with the
modified integrals
\begin{eqnarray}
{\cal Z}&=&\int\d{\bf b}\d{\bf f}\d{\bf
p}\exp\left[-\frac{\kappa}{2k_BT
\ln(4d/L)} \left(s_1 f_x^2+s_2 f_y^2\right)-i({\bf b}-{\bf
f})\cdot{\bf
p}\right]\nonumber\\
&\times&\left(1+\Gamma_x{b_x}^2+\Gamma_y{b_y}^2\right)
\left(W_0+W_{1x}p_x^2+W_{1y}p_y^2+\cdots\right) .
\end{eqnarray}
In the above equation, $\left\{W_0,\,W_{1x}, \,W_{1y},\,
\cdots\right\}$
refer to the results of the remaining integrations,
which are performed after expanding
$\exp[-k_BT\,v({\bf p},{\bf Q}^{(1)},{\bf Q}^{(2)})/2\kappa]$
in powers of ${\bf p}$, and are independent of ${\bf p}$, ${\bf b}$,
and ${\bf f}$.  After integrating over ${\bf f}$, and dropping an
unimportant constant, we obtain
\begin{eqnarray}
{\cal Z}&=&\int\d{\bf b}\d{\bf p}\;\exp\left[
-i{\bf b}\cdot{\bf p}\right]\\
&\times&\left(1+\Gamma_x{b_x}^2+\Gamma_y{b_y}^2\right)
\left[W_0-W_0\frac{k_BT\ln(4d/L)}{2\kappa}
\left(\frac{p_x^2}{s_1}+\frac{p_y^2}{s_2}\right)+\cdots\right].
\nonumber
\end{eqnarray}
Note that the $W_{1x}$ and $W_{1y}$ have been dropped since
they are subleading in the limit $d\gg L$.
Integrating over ${\bf b}$ and ${\bf p}$ then gives
\begin{equation}
{\cal Z}=
W_0+W_0\frac{k_BT\ln(4d/L)}{\kappa}
\left(\frac{\Gamma_x}{s_1}+\frac{\Gamma_y}{s_2}\right)+\cdots .
\label{b}
\end{equation}

As discussed in Sec.\ref{smemb}, we assume that the
size of the membrane is much less than the persistence length $\xi$.
Thus, the higher order terms in the expansion in Eq.(\ref{b}) are
smaller
by powers of
\begin{equation}
\frac{k_BT \ln(4d/L)}{\kappa}\approx \frac{\ln(d/L)}{\ln(\xi/a)} \ll
1.
\end{equation}
Here we have used the result\cite{DGT}
$\xi\approx a\exp\left( 2\pi\kappa/k_BT\right)$, with
a short-distance cutoff $a$ of order molecular size,
leading to the hierarchy of length scales $a<L\ll d \ll\xi$.
To leading order, then, we have ${\cal Z}=W_0$, which is independent
of $\Gamma_x$ and $\Gamma_y$ and therefore the lowest order
term in the expansion in $b$. It is interesting to note that the
above
argument does not hold for films controlled by surface tension,
as discussed in Sec.\ref{sfilm}.

\section{Solution of the biharmonic equation}\label{Abih}

The biharmonic equation (Eq.(\ref{biharm})) for a single
rod is discussed in detail in this appendix. The problem is
to find the solution to

\begin{equation}
\nabla^4 h=0 ,
\end{equation}
on a finite disk of radius $d$ from which a line segment of length
$L$
along the $y$-axis has been removed. The boundary conditions are
\begin{eqnarray}
h(x=0,-\frac{L}{2}\leq y\leq \frac{L}{2})&=&a+b_y y+g_{yy} y^2,
\label{BC2}\\
\frac{\partial}{\partial{x}}h(x=0,-\frac{L}{2}\leq y\leq
\frac{L}{2})&=&
b_x+2g_{xy} y, \nonumber\\
h(d)&=&0 ,           \nonumber\\
\frac{\partial}{\partial{r}}h(d)&=&0 \nonumber.
\end{eqnarray}
Note that for the boundary conditions, the derivatives are taken
before the limit $\epsilon \to0$. It turns out to be easier to impose
a weaker boundary condition at $r=d$, namely
\begin{equation}
 h(d)=O\left({L\over d}\right), \hskip.5cm
\frac{\partial}{\partial{r}}h(d)=O\left({L\over d}\right).\label{dbc}
\end{equation}
Since we have $d\gg L$, it will suffice to keep the leading terms in
the limit $L/d\to0$.
Performing the integration by parts yields,
\begin{eqnarray}\label{S2}
\int_{{\rm IR}^2-L}&&\d^2r (\nabla^2 h)^2               \\
&&=\int_{{\partial}({\rm IR}^2-L)} \d l \left(\nabla^2h
\frac{\partial h}
{\partial n}-h \frac{\partial\nabla^2h}{\partial n} \right)
\nonumber\\
&&=\int_{-L/2}^{L/2} \d y \left(h(0,y) f_1(y)
-\frac{\partial h}{\partial{x}}(0,y) f_2(y) \right) \nonumber,
\end{eqnarray}
where
\begin{eqnarray}
f_1(y)&=&\left.\frac{\partial\nabla^2h}{\partial x}\right|_{x=0^{+}}
-\left.\frac{\partial\nabla^2h}{\partial x}\right|_{x=0^{-}}, \\
f_2(y)&=&\left.\nabla^2 h\right|_{x=0^{+}}-
\left.\nabla^2 h\right|_{x=0^{-}}  \nonumber,
\end{eqnarray}
and, as in the text, we have denoted the finite disk of radius $d$ by
${\rm IR}^2$ for simplicity. It is easy to check that the above
boundary value problem
on ${\rm IR}^2-L$ is completely equivalent to the problem
\begin{equation}\label{C6}
\nabla^4 h= f_1(y) \delta(x)
+f_2(y) \frac{\partial}{\partial{x}}\delta(x)
\end{equation}
on ${\rm IR}^2$, provided that the conditions in Eqs.(\ref{dbc})
at $r=d$ are satisfied.
The solution to Eq.(\ref{C6}) can be given in terms of the
unknown functions $f_1(y)$ and $f_2(y)$ as
\begin{eqnarray}
h(x,y)&=&\int_{-L/2}^{L/2} \d y'\;G_1(x,y;x'=0,y')
f_1(y')\\
&+&\int_{-L/2}^{L/2} \d y'\;G_2(x,y;x'=0,y') f_2(y')  \nonumber.
\end{eqnarray}
The Green's functions $G_i$, which satisfy
\begin{eqnarray}
\nabla^4 G_1(x,y;x',y')&=&\delta(x-x')\;\delta(y-y') ,\\
\nabla^4 G_2(x,y;x',y')&=&\frac{\partial}{\partial{x}}\delta(x-x')
\;\delta(y-y') ,\nonumber
\end{eqnarray}
and obey the conditions in Eq.(\ref{dbc}) at $r=d$, are given by
\begin{eqnarray}
G_1(x,y,y')&=&\frac{1}{16\pi}\left[x^2+(y-y')^2
\right]\ln\left[\frac{x^2+(y-y')^2}{d^2}\right] \nonumber\\
&+&\frac{1}{8\pi}\frac{yy'}{d^2}(r^2+r'^2)+
\frac{1}{16\pi}[d^2-r^2-r'^2] \label{G1}, \\
G_2(x,y,y')&=&\frac{x}{8\pi}\left\{\ln\left[\frac{x^2+(y-y')^2}
{d^2}\right]+2  \frac{yy'}{d^2}+ 1-  \left(\frac{r^2+r'^2}{d^2}
\right)\right\}\nonumber.
\end{eqnarray}
Note that the boundary conditions in Eq.(\ref{dbc}) do not uniquely
specify $G_1$ and $G_2$, but allow different choices which differ by
subleading $O(L/d)$ terms at $r=d$. Indeed the asymmetry in
$G_1$ with respect to interchange of $x$ and $y-y'$ is a result
of this freedom. If we require $h$ and $\partial h/\partial r$ to
vanish at $r=d$ then $G_1$ would be rotationally symmetric.
The unknown functions $f_i$ in the above solution can now be
obtained self-consistently by matching to the known forms
of $h$ and  ${\partial h}/{\partial{x}}$ on $L$, as given by the
boundary conditions in Eq.(\ref{BC2}).
We thus end up with the following integral equations

\begin{eqnarray}
a+b_y y+g_{yy} y^2&=&\int_{-L/2}^{L/2} \d y'\;G_1(x=0,y;x'=0,y')
\;f_1(y')  \label{IQF1} \\
b_x+2 g_{xy} y&=&\int_{-L/2}^{L/2} \d y'\;
\frac{\partial}{\partial{x}}G_2(x=0,y;x'=0,y')\; f_2(y')
\label{IQF2}.
\end{eqnarray}
(Note that at $x=0$, $G_2$ and $\partial G_1/\partial x$ are both
identically zero.)
We start with Eq.(\ref{IQF2}) for $f_2(y)$, which
is somewhat easier to solve.  After changing variables to
$y=(L\cos\phi)/2$ and $y'=(L\cos\phi')/2$, this equation reads
\begin{eqnarray}
b_x+L g_{xy} \cos\phi=\frac{L}{2} \int_{0}^{\pi} \d \phi'
\sin\phi'\;f_2\left(\frac{L}{2}\cos\phi'\right)\; G'(\phi,\phi')
\label{IF2},
\end{eqnarray}
where
\begin{eqnarray}
G'(\phi,\phi')=\frac{1}{8\pi}\left[2\;\ln\left(2|\cos\phi-\cos\phi'|
\right)
-2\;\ln\left(\frac{4d}{L}\right)+1\right] \label{G'}.
\end{eqnarray}
We now use the expansion\cite{MF}
\begin{eqnarray}
\ln\left(2|\cos\phi-\cos\phi'|\right)=-\sum_{n=1}^{\infty}
{2\over n} \cos n\phi \cos n\phi' ,\label{EXP}
\end{eqnarray}
and define a series

\begin{equation}
\sin\phi' f_2\left(\frac{L}{2}\cos\phi'\right)
=\sum_{m=0}^{\infty} a_m \cos m\phi'.
\end{equation}
Solving Eq.(\ref{IF2}) for the $a_m$'s gives, to leading order in
$d\gg L$,
\begin{eqnarray}
f_2\left(\frac{L}{2}\cos\phi'\right)
=\frac{1}{\sin\phi'}\left[\frac{-8
b_x}{L\;\ln\left(\frac{4d}{L}\right)}
-8 g_{xy} \cos\phi'\right]  \label{F2}.
\end{eqnarray}

The integral equation for $f_1$ requires more care. First, note that
the
choice of $G_1$ in Eq.(\ref{G1}) does not lead to a vanishing normal
derivative at $r=d$, unless the condition
\begin{equation}
\int_{-L/2}^{L/2} \d y' y'^2 f_1(y')=0 ,
\end{equation}
is satisfied. Setting up the expansion

\begin{eqnarray}
\sin\phi' f_1\left(\frac{L}{2}\cos\phi'\right)
=\sum_{m=0}^{\infty} b_m \cos m\phi' \label{F1EXP}
\end{eqnarray}
for $f_1$, this requirement implies
\begin{eqnarray}
2b_0+b_2=0 \label{CB0B2}.
\end{eqnarray}
The integral equation (\ref{IQF1}) can now be written as
\begin{eqnarray}
a+\frac{L^2}{8} g_{yy}+\frac{L}{2}b_y \cos\phi
+\frac{L^2}{8} g_{yy} \cos2\phi
=\frac{L}{2} \int_{0}^{\pi} \d \phi' \sin\phi'
f_1\left(\frac{L}{2}\cos\phi'\right) G(\phi,\phi') \label{IF1},
\end{eqnarray}
where
\begin{eqnarray}\label{G}
G(\phi,\phi')&=&\frac{L^2}{32\pi}\left[(\cos\phi-\cos\phi')^2\;
\ln\left(2|\cos\phi-\cos\phi'|\right) \hskip
2cm\right.\nonumber\\
&&\hskip 2cm\left.-(\cos\phi-\cos\phi')^2
\;\ln\left(\frac{4d}{L}\right)
+2 \frac{d^2}{L^2}-\frac{1}{2} \cos^2\phi\right] \\
&=&-\frac{L^2}{32\pi}\left\{-2\frac{d^2}{L^2}+\ln\left(\frac{4d}{L}
\right)
-\frac{3}{4}+\left[-\frac{3}{4}+\frac{1}{2}\ln\left(\frac{4d}{L}\right
)
\right]\cos2\phi' \right.\nonumber\\
&& \hskip 1cm +\cos\phi \left[\left(\frac{5}{2}
-2\ln\left(\frac{4d}{L}\right)\right) \cos\phi'
+\frac{1}{6}\cos3\phi'\right]\nonumber\\
&& \hskip 1cm +\cos2\phi \left[-\frac{1}{2}+\frac{1}{2}
\ln\left(\frac{4d}{L}\right)-\frac{1}{3}\cos2\phi'
+\frac{1}{24}\cos4\phi'\right] \nonumber\\
&& \hskip 1cm+\sum_{n=3}^{\infty} \cos n\phi \left[\left(
\frac{2}{n}-\frac{1}{n-1}-\frac{1}{n+1}\right)\cos n\phi'\right].
\nonumber\\
&& \hskip 3cm+\frac{1}{2}\left(
\frac{1}{n+2}+\frac{1}{n}-\frac{2}{n+1}\right)\cos(n+2)\phi'\nonumber
\\
&& \left.\left.\hskip 3cm+\frac{1}{2}\left(
\frac{1}{n-2}+\frac{1}{n}-\frac{2}{n-1}\right)\cos(n-2)\phi'
\right]\right\}  \nonumber.
\end{eqnarray}
In going to the second form of $G(\phi,\phi')$ in Eq.(\ref{G}),
we have used the expansion in
Eq.(\ref{EXP}) and rearranged the resulting expression as
a series expansion so that it resembles the LHS of the integral
equation. Substituting the expansion of Eq.(\ref{F1EXP}) in the
integral
equation (\ref{IF1}) and equating the coefficients of
$\cos n\phi$ on both sides, we obtain the following set of linear
equations for the $b_n$,
\begin{eqnarray}
&(I):\quad&\left(\frac{2}{n}-\frac{1}{n-1}-\frac{1}{n+1}\right)b_n
+\frac{1}{2}\left(\frac{1}{n+2}+\frac{1}{n}-\frac{2}{n+1}\right)b_{n+2
} \\
&& \hskip 4cm
 +\frac{1}{2}\left(\frac{1}{n-2}+\frac{1}{n}-\frac{2}{n-1}\right)b_{n-2
}=0
\quad (n>2), \nonumber\\
&(II):\quad&
\left[\frac{5}{2}-2\ln\left(\frac{4d}{L}\right)\right]b_1
+\frac{1}{6}b_3=-\frac{64}{L^2} b_y,\nonumber\\
&(III):\quad& 2\left[-2\frac{d^2}{L^2}+\ln\left(\frac{4d}{L}\right)
-\frac{3}{4} \right]b_0+\left[-\frac{3}{4}+\frac{1}{2}
\ln\left(\frac{4d}{L}\right)\right]b_2=-\frac{128}{L^3}a-\frac{16}{L}
g_{yy}
\nonumber,\\
&(IV):\quad& 2
\left[-\frac{1}{2}+\frac{1}{2}\ln\left(\frac{4d}{L}\right)\right]
b_0-\frac{1}{3}b_2 +\frac{1}{24} b_4=-\frac{16}{L} g_{yy}\nonumber.
\end{eqnarray}
The solution to the above equations is (to leading order in $d\gg
L$),
\begin{eqnarray}
&&b_0=\left(\frac{L^2}{d^2}\right)\left[\frac{4}{L}g_{yy}
+\frac{32}{L^3}a\right] ,\\
&&b_1=\frac{32}{L^2\;\ln\left(\frac{4d}{L}\right)}b_y ,\nonumber\\
&&b_2=-2\left(\frac{L^2}{d^2}\right)\left[\frac{4}{L}g_{yy}
+\frac{32}{L^3}a\right],\nonumber\\
&&b_3=0,\nonumber\\
&&b_4=-\frac{384}{L}g_{yy},\nonumber
\end{eqnarray}
and all other $b_n$ are determined by the recursion relation (I).
Putting the results for $f_1$ and $f_2$ into Eq.(\ref{S2}), we find
Eq.(\ref{S}), with $s_1=s_2=4\pi$.

\section{Spherical inclusions}\label{Asph}

In this appendix we correct an error in Ref.\cite{goul}. The
expression
for ${\cal H}$ below Eq.(8) in Ref.\cite{goul} should read
\begin{equation}
{\cal H}=\frac{(k_BT)^2}{64\kappa_0A}
\left(Q^{(1)}_{ij}Q^{(1)}_{ji}+2Q^{(1)}_{ii}
+Q^{(2)}_{ij}Q^{(2)}_{ji}+2Q^{(2)}_{ii}\right)-\frac{(k_BT)^2}{2\kappa
_0}
V_1\left({\bf Q}^{(1)},{\bf Q}^{(2)}\right).
\end{equation}
This changes the final answer by a factor of 1/2. Thus, Eq.(10) of
Ref.\cite{goul} for the interaction between two inclusions of area
$A$,
separated by a distance $R$, becomes
\begin{equation}
V^T(R)=-k_BT\frac{6A^2}{\pi^2 R^4}.
\end{equation}



\begin{references}

\bibitem{alb}
 B. Alberts, J. Lewis, M. Raff, K. Roberts and J.D. Watson, {\it
Molecular
Biology of the Cell}, Garland, New York 1994.
\bibitem{gen}
 R.B. Gennis, {\it Biomembranes, Molecular Structure and Function},
 Springer-Verlag, New York 1989.
\bibitem{Las}
D.D. Lasic, {\it Liposomes from Physics to Applications}, Elsevier,
Amsterdam 1993.
\bibitem{CM}
Cevc and D. Marsh, {\it Phospholipid Bilayers: Physical Principles and
Models}, John Wiley, New York 1987.
\bibitem{Fen}
J.H. Fendler, {\it Membrane mimetic chemistry. Characterizations and
applications of micelles, microemulsions, monolayers, bilayers,
vesicles, host-guest systems, and polyions}. John Wiley, New York 1982.
\bibitem{isr}
 J. Israelachvili, {\it Intermolecular and Surface Forces}, Academic
Press, San Diego 1992.
\bibitem{mou}
 O.G. Mouritsen and M. Bloom, {\it Annu. Rev. Biophys. Biomol.
Struct.},
 {\bf 22}, 145 (1993).
\bibitem{goul}
 M. Goulian , R. Bruinsma, and P. Pincus, {\it Europhys. Lett.},
{\bf 22}, 145
(1993); Erratum in {\it Europhys. Lett.} {\bf 23}, 155 (1993).
 \bibitem{dan}
 N. Dan, P. Pincus, and S.A. Safran, {\it Langmuir}, {\bf 9}, 2768 (1993).
\bibitem{C}
 P.B. Canham, {\it J. Theor. Biol.}, {\bf 26}, 61 (1970);
\bibitem{H}
 W. Helfrich, {\it Z. Naturforsch.}, {\bf 28c}, 693 (1973).
\bibitem{DGT}
 P.G. De Gennes and C. Taupin, {\it J. Phys. Chem.}, {\bf 86}, 2294 (1982).
\bibitem{BL}
 F. Brochard and J.F. Lennon, {\it J. de Phys.}, {\bf 36}, 1035 (1975).
\bibitem{DL}
F. David and S. Leibler, {\it J. de Phys.} II, {\bf 1}, 959 (1991).
\bibitem{S}
 U. Seifert, {\it Z. Phys. B}, {\bf 97}, 299 (1995).
\bibitem{EPL}
R. Golestanian, M. Goulian, and M. Kardar, {\it Europhys. Lett.}, {\bf 33},
241 (1996).
\bibitem{ros}
 R.E. Rosensweig, {\it Ferrohydrodynamics}, Cambridge, New York 1985.
\bibitem{PL}
J.-M. Park and T.C. Lubensky, preprint.
\bibitem{LiK}
H. Li and M. Kardar, {\it Phys. Rev. Lett.} {\bf 67}, 3275 (1991);
{\it Phys. Rev.} {\bf A 46}, 6490 (1992).
\bibitem{Kozlov}
M.M. Kozlov and W. Helfrich, {\it Phys. Rev.}{\bf E 51}, 3324 (1995).
\bibitem{Podgornik}
R. Podgornik, preprint, cond-mat/9503146 (1995).
\bibitem{ramin}
R. Golestanian, submitted to Europhys. Lett. (1996).
\bibitem{MF}
 P.M. Morse, H. Feshbach, {\it Methods Of Theoretical Physics},
McGraw-Hill,
New York 1953.

\end{references}
\end{document}